\documentclass[showpacs,twocolumn,nofootinbib,superscriptaddress,floatfix]{revtex4-1}
\usepackage{amsfonts,amsmath,amssymb,amsthm}
\usepackage[bookmarks, bookmarksopen, bookmarksnumbered]{hyperref}

\usepackage{graphicx}
\usepackage{color}
\usepackage[utf8]{inputenc}
\usepackage[T1]{fontenc}
\usepackage{lmodern}
\usepackage[right]{rotating}
\usepackage{longtable}
\usepackage{array}
\usepackage{multirow}
\usepackage{paralist}
\usepackage{siunitx}
\usepackage{physics}

\newcommand{\codename}[1]{\texttt{#1}}

\graphicspath{{pdf/}{png/}}

\newcommand{\tmerger}{t_\mathrm{merger}}

\begin{document}
\title{Modeling Mergers of Known Galactic Systems of Binary Neutron Stars}
\date{\today}

\author{Alessandra \surname{Feo}}
\affiliation{Parma University and INFN Parma, Parco Area delle Scienze 7/A, I-43124 Parma (PR), Italy}
\author{Roberto \surname{De Pietri}}
\affiliation{Parma University and INFN Parma, Parco Area delle Scienze 7/A, I-43124 Parma (PR), Italy}
\author{Francesco  \surname{Maione}}
\affiliation{Parma University and INFN Parma, Parco Area delle Scienze 7/A, I-43124 Parma (PR), Italy}
\author{Frank \surname{L\"offler}}
\affiliation{Center for Computation \& Technology, Louisiana State University, Baton Rouge, LA 70803 USA}

\begin{abstract}
We present a study of the merger of six different known galactic systems of
binary neutron stars (BNS) of unequal mass with a mass ratio between $0.75$ and
$0.99$. Specifically, these systems are J1756-2251, J0737-3039A, J1906+0746,
B1534+12, J0453+1559 and B1913+16.
We follow the dynamics of the merger from the late stage of the inspiral
process up to $\sim$\SI{20}{ms} after the system has merged, either to form a
hyper-massive neutron star (NS) or a rotating black hole (BH), using a
semi-realistic equation of state (EOS), namely the seven-segment piece-wise
polytropic SLy with a thermal component. For the most extreme of these
systems ($q=0.75$, J0453+1559), we also investigate the effects of different
EOSs: APR4, H4, and MS1.
Our numerical simulations are performed using only publicly available open
source code such as, the Einstein Toolkit code deployed for the dynamical
evolution and the LORENE code for the generation of the initial models.
We show results on the gravitational wave signals, spectrogram and 
frequencies of the BNS after the merger and the BH properties in the two 
cases in which the system collapse within the simulated time.
\end{abstract}

\LTcapwidth=\columnwidth

\pacs{
04.25.D-,  
04.40.Dg,  
95.30.Lz,  
97.60.Jd   
}

\maketitle

\section{Introduction}
\label{sec:intro}
The direct observation of the gravitational wave (GW) signal in events GW150914 
and GW151226 \cite{FirstDetection,Abbott:2016nmj}, emitted by the coalescence 
of a compact binary system composed by two black holes (BH) 
is the dawn of the GW astronomy era. The measured signal from those events shows
the expected signature of an inspiral and merger, and subsequent 
final black hole ringdown. This single event encourages further studies of properties
and astrophysical implications of binary black hole (BBH) populations
\cite{TheLIGOScientific:2016wfe,TheLIGOScientific:2016htt}. 
Among compact binary systems, binary neutron star (BNS) mergers are 
another expected source candidate for GW astronomy.
Their composition and ultra-dense neutron star (NS) matter behavior encoded 
in the Equation of State (EOS) make them a unique target to be measured with 
ground-based interferometers (Advanced LIGO \cite{TheLIGOScientific:2014jea}, 
Advanced Virgo \cite{TheVirgo:2014hva}, and KAGRA \cite{Aso:2013eba}),
in order to understand the structure and properties of NS systems and the 
distribution of NS masses.

The first BNS system discovered is PSR B1913+16,
better known as Hulse-Taylor binary \cite{Hulse:1974eb}, 
a pulsar which, together with another NS is in orbit around a common
center of mass forming a binary star system. This system 
provided indirect evidence of the emission of GW as predicted by General 
Relativity~\cite{Hulse:1974eb,Weisberg:2010zz}
(from now on we will use the pulsar name to denote the 
corresponding BNS system).
Since then, nine other binary neutron star systems have been 
discovered in our galaxy (see Ref.~\cite{Martinez:2015mya} for details).
Among them is also the binary system of the pulsar 
J0453+1559~\cite{Martinez:2015mya}, with a mass ratio of 
$(q=M_1/M_2=0.75)$ \cite{Martinez:2015mya},
confirming the existence of unequal mass BNS systems 
in nature with a quite unexpected large mass difference between the 
two stars. This fact has the potential to change and improve our 
understanding on the nature of NS formation as well as their evolution 
and final fate. Because of this, it also should be considered when computing
templates of GW signals from BNS mergers for data analysis pipelines.

\newcommand{\UOmega}{($\mathrm{krad}/\mathrm{s}$)}
\newcommand{\UnitJ}{($GM_\odot^2/c$)}
\newcommand{\UnitM}{($M_\odot$)}

\begin{table*}
\begin{tabular}{l|llccrc|cccccc|l}
Pulsar  & $M_p$  & $M_c$  & $q$ & e & $\tmerger$ &e${}_{10}$&$M^{(1)}$ &$M^{(2)}$ &e${}_\mathrm{ID}$& $\Omega$ & $M_\mathrm{ADM}$  & $J$    & Reference\\
        & \UnitM & \UnitM &   &   & (Gyr)      & at 10 Hz &\UnitM    &\UnitM    &          & \UOmega  & \UnitM            & \UnitJ & \\
\hline
J1756-2251  &1.341(7)  &1.230(7)  &0.92 & 0.18056 & 15.85 & $7.20 \times 10^{-7}$ &1.33 &1.47 & 0.022 & 1.773 & 2.548 & 6.654 &\cite{Faulkner:2004ha}\\
J0737-3039A &1.3381(7) &1.2489(7) &0.93 & 0.08778 & 0.08  & $1.11 \times 10^{-6}$ &1.36 &1.47 & 0.023 & 1.777 & 2.564 & 6.728 &\cite{Burgay:2003jj,Kramer:2006nb}\\
J1906+0746  &1.291(11) &1.322(11) &0.98 & 0.08530 & 0.32  & $6.48 \times 10^{-7}$ &1.41 &1.45 & 0.024 & 1.784 & 2.589 & 6.848 &\cite{Lorimer:2005un,vanLeeuwen:2014sca}\\
B1534+12    &1.3330(2) &1.3454(2) &0.99 & 0.27368 & 2.51  & $8.85 \times 10^{-8}$ &1.46 &1.47 & 0.025 & 1.801 & 2.653 & 7.136 &\cite{Wolszczan:1991kj,Fonseca:2014qla}\\
J0453+1559  &1.559(5)  &1.174(4)  &0.75 & 0.11252 & 14.55 & $1.14 \times 10^{-8}$ &1.27 &1.74 & 0.030 & 1.816 & 2.708 & 7.238 &\cite{Martinez:2015mya}\\
B1913+16    &1.4398(2) &1.3886(2) &0.96 & 0.61713 & 0.32  & $5.32 \times 10^{-6}$ &1.53 &1.59 & 0.025 & 1.840 & 2.801 & 7.816 &\cite{Hulse:1974eb,Weisberg:2010zz}\\
\end{tabular}
\caption{Known BNSs in our galaxy, as appearing in 
Ref.\cite{Martinez:2015mya}. $M_p$ is the mass of the pulsar that denotes the system 
and $M_c$ the mass of its companion, $q$ represents the mass ratio, and $e$ the eccentricity.
$t_\mathrm{merger}$ is the merger time of the system computed using Eq.~(7) of~\cite{lrr-2008-8},
and e${}_{10}$ is the eccentricity when the rotation frequency of the system 
is 10~Hz, computed using Eq.~(1.2) of~\cite{Moore:2016qxz}.
The remaining columns represent the properties of the initial 
data used to model the BNS system using the SLy EOS, where $M^{(1)}$ and $M^{(2)}$ are the 
baryonic masses, e${}_\mathrm{ID}$ is the measured eccentricity, $\Omega$ the initial
rotation frequency, 
$M_\mathrm{ADM}$ the ADM mass of the system and $J$ its angular momentum.   
\label{table:galaxy}}
\end{table*}

In this paper we analyze all known BNS systems in our galaxy
for which both masses are known (see Table~\ref{table:galaxy}
\cite{Martinez:2015mya}). They contain both the recently
discovered system J0453+1559 ($q=0.75$)~\cite{Martinez:2015mya},
as well as J0737-3039A, which is the most relativistic binary 
BNS known today~\cite{Burgay:2003jj,Kramer:2006nb} and the
only binary pulsar.
Driven by the experimental discovery of a $q=0.75$ BNS,
we present three-dimensional numerical simulations of the 
dynamics of BNS mergers with different total
baryonic mass and different mass ratio up to $q=0.75$, to understand the 
impact of the mass ratio on the GW signal emitted in both the last orbits
before merger and in the post-merger phase.
For all systems, we use the SLy EOS~\cite{Douchin01}, based on the 
Skyrme Lyon effective nuclear interaction, where a 
semi-realistic seven-segment piece-wise (isentropic)-polytropic approximant 
\cite{Read:2009constraints} is implemented, namely the SLyPP EOS,
including a thermal component given by $\Gamma_{th} = 1.8$. 
We use the BSSN-NOK~\cite{Nakamura:1987zz,Shibata:1995we,Baumgarte:1998te,Alcubierre:2000xu,Alcubierre:2002kk} method for the evolution
of the gravitational sector and WENO~\cite{shu:98} as a reconstruction method for the 
matter sector. This study has been performed using only publicly
available open source software, in particular, the Einstein
Toolkit (ET)~\cite{Loffler:2011ay,EinsteinToolkit:web},
used for the dynamical evolution, and the LORENE code~\cite{lorene:web,Gourgoulhon:2000nn}
to generate the initial models, similar to studies
recently published in Refs.~\cite{DePietri:2015lya,Maione:2016zqz}. 

For the most extreme model, the J0453+1559 system
($q=0.75$), we also investigate the effects of using different EOS for
nuclear matter, which include: the APR4 EOS \cite{Akmal:1998cf},
the H4 EOS \cite{Lackey:2005tk} and the MS1 EOS \cite{Muller:1995ji}.

The organization of the paper is as follows. In Sec.~\ref{sec:setup} we briefly
review the numerical setup used. A detailed discussion of an essentially
identical setup can be found in~\cite{DePietri:2015lya}). In
Sec.~\ref{sec:results}, we discuss each of the models and present the results of
our simulations: the fate of these systems, the properties of the remnants, and
the effect of the EOS for the case of J0453+1559. A summary and conclusions are
given in Sec.~\ref{sec:conclusions}.

In this work we use a space-like signature $-,+,+,+$, with Greek indices 
running from 0 to 3, Latin indices from 1 to 3, and the standard convention for summation
over repeated indices.  The computations are performed using the standard $3+1$ split
into (usually) space-like coordinates $(x,y,z)=x^i$ and a time-like coordinate $t$.
Our coordinate system $(x^\mu)=(t,x^i)=(t,x,y,z)$ (far-from the origin)  
are, as it can be checked, almost isotropic coordinates and  (far-from the origin) 
they would have the usual  measure unit of ``time'' and ``space'' and in particular $t$ 
is close to be identified as the time measured from an observer at infinity.
All computations are performed in normalized computational units 
(hereafter denoted as CU) in which $c=G=M_\odot=1$. We report the radius of the 
sphere used for gravitational waves extraction in CUs.

\section{Numerical Setup}
\label{sec:setup}
The systems analyzed in this paper share a similar dynamics pre-merger: an
inspiral driven by gravitational wave radiation. We follow the remnant of the
merger for approximately $20$~ms post-merger to determine its properties, and
to be able to measure gravitational waves leaving the system. As will be shown
in Sec.~\ref{sec:results}, some of these collapse to a BH,
requiring a black hole horizon finder.

We have analyzed similar models in the past for ad-hoc system parameters. The
same numerical setup can be used for the current study, which is why we will
only briefly summarize the setup here and refer the reader
to Ref.~\cite{DePietri:2015lya} for details. One fundamental aspect of the present
study is that all data has been produced using only freely available, open
source software. Moreover, all information necessary to reproduce and
re-analyze our simulations has been made available in the same way.
Detailed instruction on how to achieve this can be found on the 
Subversion server of the gravity group at Parma University.

In order to describe BNS systems we need to
use the Einstein's general relativity equations to describe the metric 
$g_{\mu \nu}$ of the dynamical spacetime
\begin{align}
R_{\mu\nu} -\frac{1}{2} g_{\mu\nu} R = & \frac{8\pi G}{c^4} T_{\mu\nu} \, .
\end{align}
The dynamical evolution is described using the Einstein Toolkit, which is a 
publicly available, community-driven general relativistic code.
In particular, we have chosen the eleventh release 
(code name ``Hilbert'', ET\_2015\_05).
The ET is based on the Cactus computational toolkit 
\cite{Cactuscode:web,Goodale:2002a,CactusUsersGuide:web}, a software
for high performance computing that uses: 
the adaptive mesh refinement (AMR) methods implemented by Carpet 
\cite{Schnetter:2003rb, Schnetter:2006pg,CarpetCode:web}.
In particular, the initial data is discretized on a Cartesian grid with  
6 levels of mesh refinement. The inner level contains a grid with a 
spacing of $dx=0.25$ CU that corresponds to $\simeq 369$ m.
We also use a mirror symmetry across the (x,y) plane
which reduces the computational cost of the simulations by a factor of 2.
The evolution of the spacetime metric is managed using the McLachlan package
\cite{McLachlan:web}, and more specifically the version implementing the
Einstein equations through 
a $3+1$ dimensional split using the BSSN-NOK formalism, 
where we use a  fourth order finite difference stencils for the curvature 
evolution and a $\Gamma$-driver shift
condition~\cite{Alcubierre:2002kk}.  During evolution, a Sommerfeld-type
radiative boundary condition is applied to all components of the evolved
BSSN-NOK variables as described in~\cite{Alcubierre:2000xu}, and the HLLE 
(Harten-Lax-van Leer-Einfeldt) approximate Riemann 
solver~\cite{Harten:1983on,Einfeldt:1988og} is used.
The initial data of our simulations are generated using the 
LORENE code~\cite{lorene:web} and in particular, we use the irrotational
BNS data generation of \cite{Gourgoulhon:2000nn}.

We also need a proper description of matter 
(details can be found in Ref.~\cite{DePietri:2015lya}). In particular, we use  
the energy-momentum tensor $T_{\mu \nu}$ to be that of an ideal relativistic fluid:
$T^{\mu \nu} = \rho (1 + \epsilon + \frac{p}{\rho})u^\mu u^\nu + P g^{\mu \nu}$, 
where $\rho$ is the rest mass density, $\epsilon$ is the specific internal matter energy,
$u^\mu$ is the 4-velocity of the matter and $P$ is the pressure.
We also use the conservation laws 
for the energy-momentum tensor, $\nabla_\mu T^{\mu\nu} = 0$ and the 
baryon density $\nabla_\mu(\rho u^\mu) = 0$, closed by an EOS of the 
type $P=P(\rho,\epsilon)$.

\begin{figure}
\begin{centering}
  \includegraphics[width=0.45\textwidth]{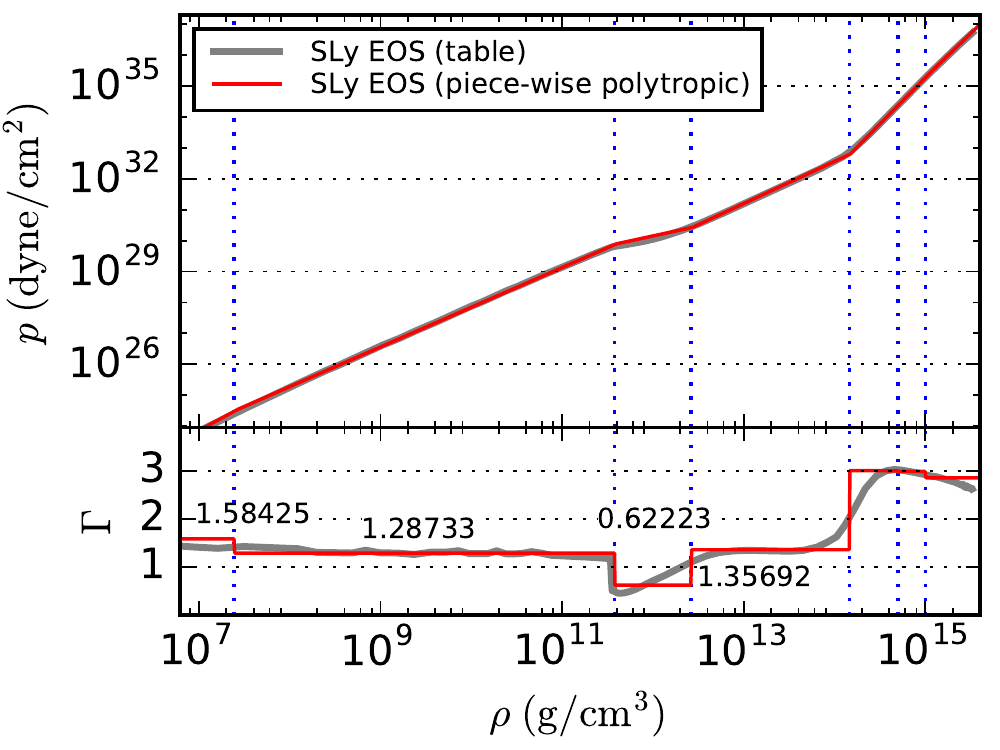}\\
\end{centering}
\vspace{-2mm}
\caption{Plot of the pressure ($p$) and of the adiabatic index ($\Gamma=d\log(p)/d\log(\rho)$) as 
a function of the baryon density ($\rho$) for the SLy EOS (tabulated) and its 
piece-wise polytropic approximation (the one used in the present work).}
\label{fig:EOS}
\end{figure}

\begin{figure}
\begin{centering}
   \includegraphics[width=0.45\textwidth]{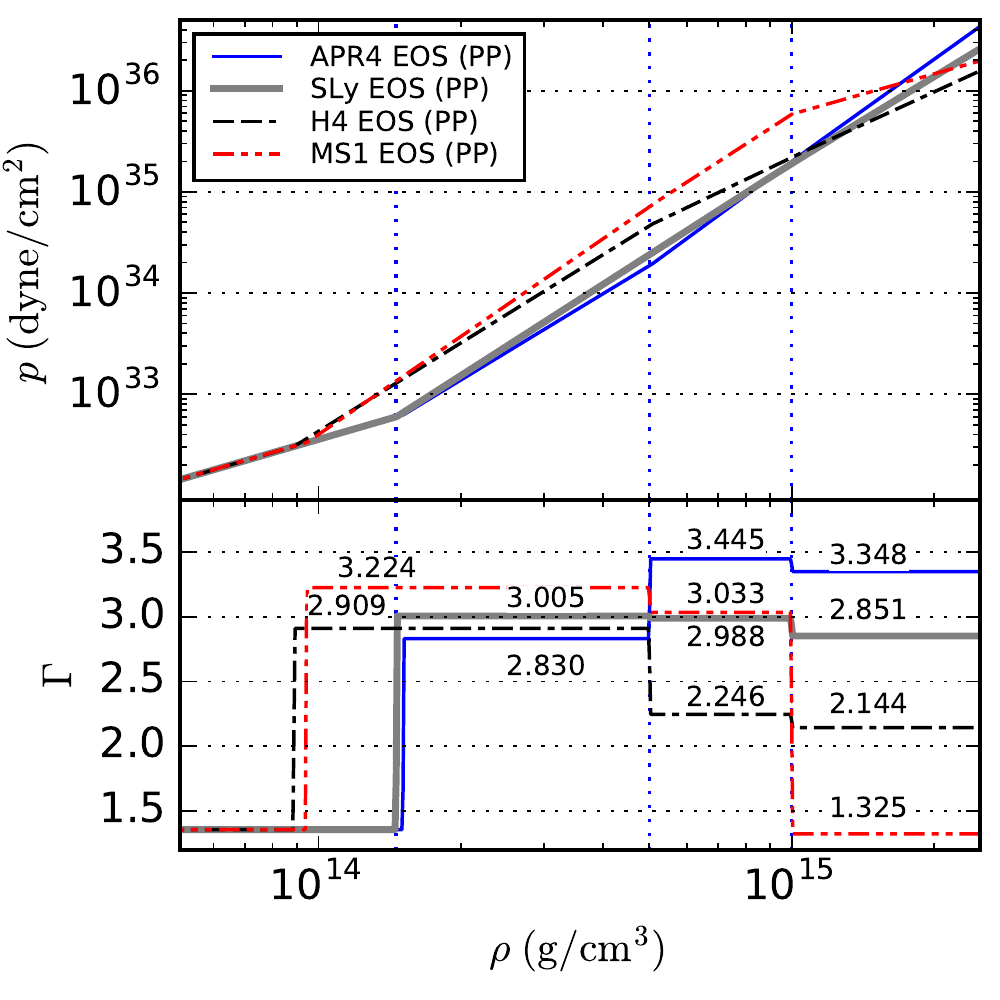}\\
\end{centering}
\vspace{-2mm}
\caption{Plot of the pressure ($p$) and of the adiabatic index ($\Gamma=d\log(p)/d\log(\rho)$) as 
a function of the baryon density ($\rho$) for the SLy EOS piece-wise polytropic 
approximation (the same used in Fig.\ref{fig:EOS}) together with the APR4, H4 and MS1 EOSs.}
\label{fig:4EOS}
\end{figure}

For all the BNS models investigated in this paper, we use, unless otherwise stated, a seven-segment 
isentropic polytropic
approximant, that we refer as SLyPP, and that was already used 
in our previous work \cite{DePietri:2015lya}. This EOS belongs to the 
SLy EOS prescription~\cite{Douchin01}, supplemented
by a thermal component $\Gamma_\textrm{th} = 1.8$ (details can be 
found in \cite{DePietri:2015lya}). In particular, we use 
four pieces for the crust and three pieces for the core~\cite{Read:2009constraints}.
The EOS is described through the expression
\begin{equation}
P(\rho,\epsilon) = P_{\mathrm{cold}}(\rho) + P_{\mathrm{th}}(\rho,\epsilon) \, ,
\end{equation}
where each density region, $\rho_i \leq \rho < \rho_{i+1}$, satisfy:
\begin{align}
P_{\mathrm{cold}} = & K_i \rho^{\Gamma_i} \, , \\  
\epsilon_{\mathrm{cold}} = & \epsilon_i + \frac{K_i}{\Gamma_i-1}\rho^{\Gamma_i-1}  
\end{align}  
and $\epsilon_i$ and the polytropic constant $K_i$ are chosen to guarantee 
the pressure and specific energy density continuity.

During the evolution, the EOS of the cold nuclear matter is supplemented by a 
thermal component of the 
form:
\begin{equation}
P_{\mathrm{th}} = \Gamma_{\mathrm{th}} \rho (\epsilon - \epsilon_{\mathrm{cold}}) 
\end{equation}
and choosing $\Gamma_{\mathrm{th}}$, as in our previous work \cite{DePietri:2015lya,Maione:2016zqz}, 
to be 1.8 according to 
\cite{bauswein:2010testing,hotokezaka:2013mass,Kyutoku:2014yba,hotokezaka:2013remnant}.
Fig.~\ref{fig:EOS} shows a plot of the pressure $p$ and the adiabatic index
$\Gamma=\dv{\log(p)}{\log(\rho)}$ as a function of the baryonic density $\rho$ 
for the SLy EOS (gray line) \cite{Douchin01} and its piece-wise polytropic 
SLy EOS (red line) \cite{Read:2009constraints}.

For the more extreme system ($q=0.75$, J0453+1559), beside the 
study of the SLy EOS, we also
investigate the effects of assuming other cold EOSs for nuclear
matter at beta equilibrium. Summarizing, we analyzed in decreasing order 
of compactness:
\begin{itemize}
\item The APR4 EOS \cite{Akmal:1998cf}, obtained using variational chain summation 
methods using the Argonne two-nucleon interaction and also including boost 
corrections and three-nucleon interactions;
\item The SLy EOS \cite{Douchin01}, based on the Skyrme Lyon effective 
nuclear interactions;
\item The H4 EOS \cite{Lackey:2005tk}, constructed in a relativistic mean field 
framework including also Hyperons contributions and tuning the parameters to
have the possible stiffest EOS compatible with astrophysical data;
\item The MS1 EOS \cite{Muller:1995ji}, constructed with relativistic means field 
theory considering only standard nuclear matter. 
\end{itemize}

We show in Fig.~\ref{fig:4EOS} 
the core part of Fig.~\ref{fig:EOS} but for all the different
EOS used, namely: APR4, H4, MS1 and the SLy EOSs.
Notice, that up to a baryonic density of the order of \SI{e14}{\g\per\cm\cubed},
which represents the separation between the crust and the core of the NS, 
we use the same prescription: the one coming from the 
prescription of the crust given in Ref.~\cite{Douchin01}
using the parametrization of Ref.~\cite{Read:2009constraints}.

\begin{figure*}
\begin{centering}
  \includegraphics[width=0.90\textwidth]{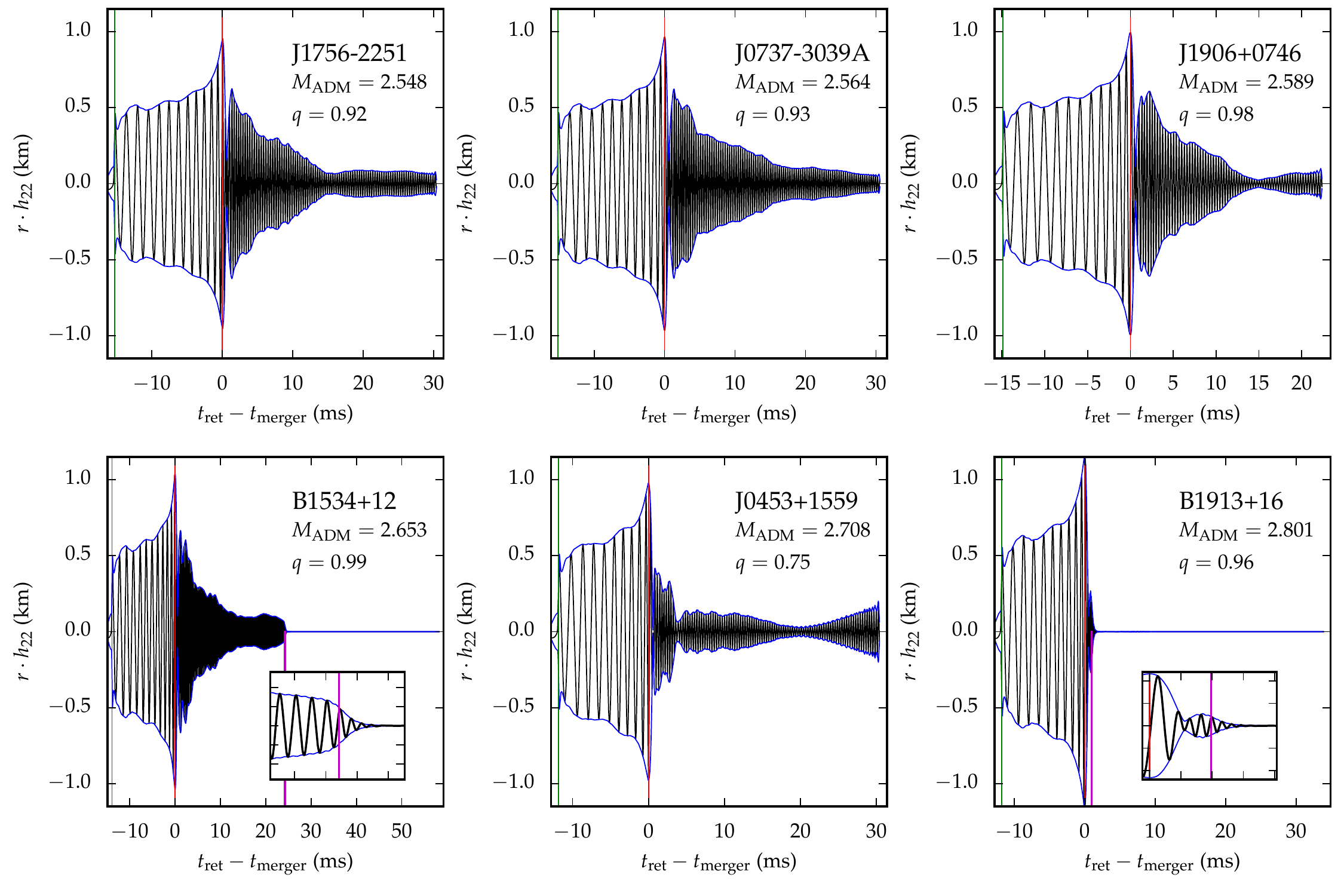}\\
  \includegraphics[width=0.90\textwidth]{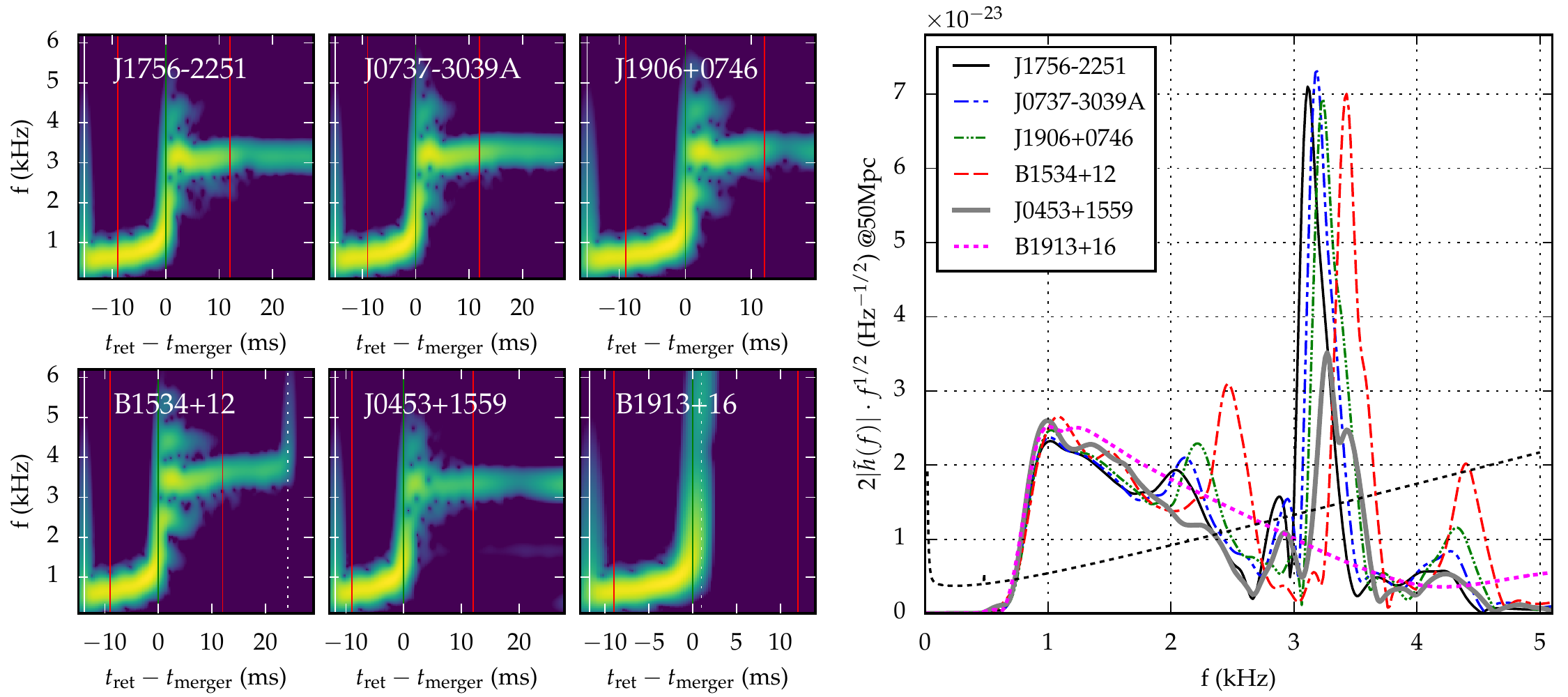}
\end{centering}
\vspace{-2mm}
\caption{Overview of the models studied using the SLy EOS. These models are 
J0453+1559, J1756-2251, J0737-3039A, B1913+16, J1906+0746, B1534+12,
with unequal mass ratios of $q=0.75,0.92,0.93,0.96,0.98,0.99 $, respectively.
We report in the top panel the amplitudes of the main $l=2, m=2$ mode
of the gravitational wave signal, and in the bottom-left panel the corresponding
spectrogram. In the bottom-right panel we show the corresponding power spectral
density for the six systems at a distance of \SI{50}{Mpc}, where the Fourier 
transform is computed on the interval going from \SI{9}{ms} before merger to \SI{12}{ms} after
merger.}
\label{fig:all}
\end{figure*}

\section{Results}
\label{sec:results}
We have analyzed the dynamics of the merger of six different BNS 
systems, that appear in Table~\ref{table:galaxy}, from the late-inspiral phase 
(last five-six orbits, depending on the model) until approximately \SI{20}{ms} after
merger.
These models are J0453+1559, J1756-2251, J0737-3039A, B1913+16, J1906+0746, B1534+12, 
with an unequal mass ratio of $q=0.75,0.92,0.93,0.96,0.98,0.99$, respectively.
In the following, we present the general characteristics of the evolution 
of these systems. 

We extracted the gravitational wave (GW) signal from each simulation using the 
same procedure of Ref.~\cite{Maione:2016zqz}. In brief, for each simulation 
we extracted the GW signal from the curvature $\psi_4$ \cite{Newman:1961qr,Baker:2001sf},
written in terms of spin-weighted spherical harmonics of 
spin $-2$ \cite{Thorne:1980ru}:
\begin{equation}
 \psi_4(t, r, \theta, \phi)  = 
 \sum_{l=2}^{\infty}{\sum_{m=-l}^{l}{\psi_4^{lm}(t,r) {{}_{-2}\,}{Y}_{lm}(\theta,\phi)}} 
\end{equation}
up to $l=6$. At null infinity, $\psi_4$ is related to the GW strain 
through the relation
\begin{equation} 
\psi_4\ =\ \ddot{h}_+ - i \ddot{h}_x := \ddot{\bar{h}} \, ,  
\label{eq:h}
\end{equation}
where $\bar{h}$ is the complex conjugation of the GW strain.
To get the GW strain components $h^{lm}$, we numerically integrate this 
expression twice 
\begin{align}
\bar{h}^{(0)}_{lm} = & \int_{0}^t{dt' \int_{0}^{t'}{dt'' \psi_4^{lm}(t'',r)}} 
\label{eq:time_int} \\
\bar{h}_{lm} = & \, \, \bar{h}^{(0)}_{lm} - Q_1 t - Q_0 
\label{eq:fit_int}
\end{align}
and we fix the two initial values of the integration procedure ($Q_0$ and $Q_1$) 
by minimizing the overall drifting. However, it is well known that this procedure still shows
low-frequency spurious oscillations in the strain amplitude. As discussed in 
\cite{Maione:2016zqz}, we remove this (quite likely due to noise aliasing problems)
by applying, after the integration and the subtraction of the linear drift, 
a digital high-pass Butterworth filter (an IIR filter) imposing a maximum signal
suppression of \SI{-0.01}{dB} at the minimum physical frequency $\hat{f}_0=\Omega/\pi$ (computed as twice
the initial orbital frequency) and a signal suppression of \SI{-80}{dB} at $\hat{f}_0/10$.

In this work we extracted the GW signal at the coordinate radius of $R=700$ 
CU (\SI{1034}{\km}) and we always report the retarded time (for GW extraction quantities) that 
corresponds to the coordinate time for quantities integrated over the numerical grid.
In detail, we have $t_\mathrm{ret}=t - R^*$, ($R^* = R + 2 M_\mathrm{ADM} \log(-1+R/2M_\mathrm{ADM})$)
since our numerical relativity coordinate system far away form the center
is almost identical to isotropic coordinates. We also implement a 1st order extrapolation 
in $1/R$, as proposed in~\cite{Lousto:2010qx} (see~\cite{Maione:2016zqz} for details).

To obtain the energy and the angular momentum budgets of the collapsed 
models we used the energy $E_\mathrm{gw}$ and the angular momentum $J_\mathrm{gw}$
carried away by GW, which are given by
\begin{align}
\frac{E_\mathrm{gw}}{dt} = & \frac{R^2}{16 \pi} \int{ d\Omega \left| \dot{h}(t,\theta,\phi) \right|^2}\, , \\ 
\frac{dJ_z^\mathrm{gw}}{dt} = & \frac{R^2}{16 \pi} Re \left[ \int{ d\Omega \left( \partial_{\phi}\; 
\dot{\bar{h}}(t,\theta,\phi) \right) h(t,\theta,\phi)  } \right]\ ,
\end{align}
as well as the mass $M_\mathrm{bh}$ and angular momentum of the BH $J_\mathrm{bh}$, 
 computed using the isolated horizon formalism~\cite{Ashtekar:2000sz,Ashtekar:2001jb,Ashtekar:2004cn} on the apparent horizon, utilizing the ET modules \codename{QuasilocalMeasures}~\cite{Dreyer:2002mx} and \codename{AHFinderDirect}~\cite{Thornburg:2003sf}. The mass and angular momentum
contribution of matter is defined in \cite{DePietri:2015lya}.

\subsection{Properties of the six galactic binary neutron star systems}
\label{SEC:fate}
The overview of the evolution of known Galactic systems of BNS is presented in 
Fig.\ref{fig:all}, where we show the amplitude of the $r\cdot h_{22}$ mode 
as a function of $(t - t_\mathrm{merger})$ (top), the spectrum of frequencies 
(spectrogram) using \SI{5}{ms} bind as a function of the time (on the bottom-left panel),
and the power spectral density (PSD, Fourier Transform) of the effective 
GW $\bar{h}(f)$, in the optimal oriented case for a source at \SI{50}{Mpc}, where
we consider the signal from \SI{9}{ms} before, to \SI{12}{ms} after merger in
the bottom-right panel.

The evolution of the six galactic systems on top of Fig.\ref{fig:all} is shown in the order of
increasing ADM mass. All these models used the SLy EOS and an interbinary distance
of \SI{44.3}{km}. All the numerical simulations presented in this paper were 
performed with a resolution of $dx=0.25$ CU which corresponds to 
$\simeq 369$ m).
The green line in each BNS represents $t_\mathrm{ret}=0$.
The red line indicates the merger time at $t=0$ ms.
We define merger time here as the time of maximal GW signal amplitude.

An additional purple line shows the formation
of a BH and only appears for two of these systems (B1913+16 and B1534+12), 
at least during the simulation time we have investigated.
These two pulsars show-case two different merger remnants involving a BH.
B1913+16 (the Hulse-Taylor binary) 
rapidly collapses to a BH, while B1534+12 shows a delayed collapse to a BH 
a few milliseconds after the merger (around \SI{25}{ms} in this particular simulation).
For both of these systems, we increased the simulation time up to \SI{40}{ms} after merger
to analyze the formation of an accretion disk (see section~\ref{SEC:BH} for details). 

Model J0453+1559, with an ADM mass in between the masses of B1534+12 and B1913+16,
does not collapse to a BH during the simulated time (around \SI{30}{ms} after
merger). This is probably related to the fact that it corresponds to an extreme 
unequal mass system (with $q=0.75$), while the other two pulsars, that do collapse to a BH,
have almost equal masses. The remaining models (see Table~\ref{table:galaxy})
form a remnant lasting more than \SI{20}{ms} after merger. All of them will 
eventually collapse to a BH since their masses are greater than 
the maximum mass that can be supported by the SLy EOS for a uniformly rotating
star. We did not follow up on those stages in this paper.

The PSD of the effective GW 
signal in Fig.\ref{fig:all} (bottom-right panel) shows 
the presence of a dominating peak for each model, except for the one 
that rapidly collapses to a BH (B1913+16).
This peak corresponds to the frequency $f_p$ (also called $f_2$ or 
$f_\mathrm{peak}$) of the fundamental quadrupolar $m=2$ oscillation mode of 
the massive NS formed after the merger \cite{stergioulas:2011gravitational},
and depends on the compactness of the star~\cite{hotokezaka:2013mass,bauswein:2014revealing,bauswein:2012measuring,bauswein:2015exploring,hotokezaka:2013remnant}.
This dependence can be seen in our data as increase of $f_p$ for more compact systems,
within a range of \SI{3}{kHz} to \SI{4}{kHz} for models
J1756-2251, J0737-3039A, J1906+0746 and B1534+12,
where B1534+12 has the largest ADM mass. The only exception is
B1913+16, whose PSD rapidly decays to zero as the pulsar 
collapses to a BH soon after the merger.
Also, J0453+1559 does not follow this trend due to its
very low $q=0.75$ value, which renders this system quite different from
the others.

Some of the models show other, secondary post-mergers peaks at frequencies $f_{-}$
and $f_{+}$ (also known as $f_1$ and $f_3$ in the literature)
that can be seen in Fig.\ref{fig:all}. They are also recognizable
from the spectrum and may help to extract NS parameters 
(radius, mass) from GW detections~\cite{Takami:2014tva,Takami:2015gxa}. 
In Table~\ref{TAB:peakFrequencies} we report the frequencies of 
all recognizable spectral peaks for the BNS systems of Fig.~\ref{fig:all}.

\begin{table}
\begin{tabular}{lcccccc}
\multirow{2}{*}{Model}        & $\tau_0$ & $f_0$ & $f_p$ &$\hat{f}_p$ & $f_-$ & $f_+$  \\
             & (ms)     & (kHz) & (kHz) &   (kHz)  & (kHz) & (kHz)  \\
\hline
J1756-2251   & 2.49 & 1.148  & 3.163 &  3.114   & 2.028 & 4.280  \\
J0737-3039A  & 2.74 & 1.139  & 3.293 &  3.182   & 2.105 & 4.271  \\
J1906+0746   & 2.72 & 1.119  & 3.326 &  3.231   & 2.206 & 4.321  \\ 
B1534+12     & 3.72 & 0.984  & 3.667 &  3.424   & 2.459 & 4.397  \\
J0453+1559   & 2.54 & 0.998  & 3.331 &  3.268   & ---  &  --- \\ 
\end{tabular}
\caption{Main peak frequencies and damping times of the post-merger
phase of the simulated models at $dx=0.25$ CU. 
$\tau_0$ and $f_0$ are the dumping time and the frequency of 
the oscillation of the lapse $\alpha$ between \num{1.5} to \SI{8}{ms} after $t_\mathrm{merger}$,
respectively.
The frequency $f_p$ is determined by a single-frequency fitting
procedure on the last part of $h_{22}$ signal.
The frequencies, $\hat{f}_p$, $f_-$, $f_+$, are derived by an analysis of 
the Fourier spectrum and represent the frequencies of the main and secondary peaks
of the PSD (bottom-right panel of Fig.~\ref{fig:all}). 
Notice that we do not include the sixth galactic system
(B1913+16) as it rapidly goes to a BH. }
\label{TAB:peakFrequencies}
\end{table}

One of the systems clearly exhibiting these three peaks is B1534+12,
with the middle peak $f_p$ and two secondary peaks in both sides of the principal frequency.
The same three peaks can be observed for models
J1756-2251, J0737-3039A and J1909+0746. See Table~\ref{TAB:peakFrequencies} for their specific
frequencies.
It should be noted that the secondary peaks $f_\pm$ decays within few \si{ms} 
as it can be seen in the spectrograms (bottom-left) of Fig.~\ref{fig:all}.

On the other hand, this three-peak-structure does not show up in the case
of the J0453+1559 ($q=0.75$). This is 
consistent with the results of Ref.~\cite{DePietri:2015lya}, where it has been 
shown that the three-peak-structure is gradually suppressed for unequal mass
BNS systems. A similar behavior (suppression of peaks) can be observed
analyzing the same system using another EOS (see Section~\ref{SEC:q=0.75} for more details).
Moreover, for the J0453+1559 system it should be noted that there is
a small shift of frequency, \SI{22}{ms} after the merger, and that the mode is growing with 
a characteristic exponential growth time of $\tau=5.75$~ms and a 
frequency equal to $f_p=3.331$~kHz. 
This growing amplitude suggests the activation of an unstable mode.
A similar phenomenon is also present for the 
J1906+0746 system, where \SI{15}{ms} after the merger a growing 
mode with $f_p=3.329$ kHz and $\tau=5.30$ ms is observed.
On the other hand, growing modes are not present for 
J1756-2251 and 0737-3039A. The presence of such 
instabilities of the remnant should not be considered unexpected,
since it is well known that differentially rotating stars show dynamical 
instabilities, and that the threshold for their activation depends on the EOS
(see~\cite{Watts:2003nn,Baiotti:2006wn,Manca:2007ca,DePietri:2014mea,Loffler:2014jma} and 
reference therein).

\begin{figure}
\begin{centering}
  \includegraphics[width=0.45\textwidth]{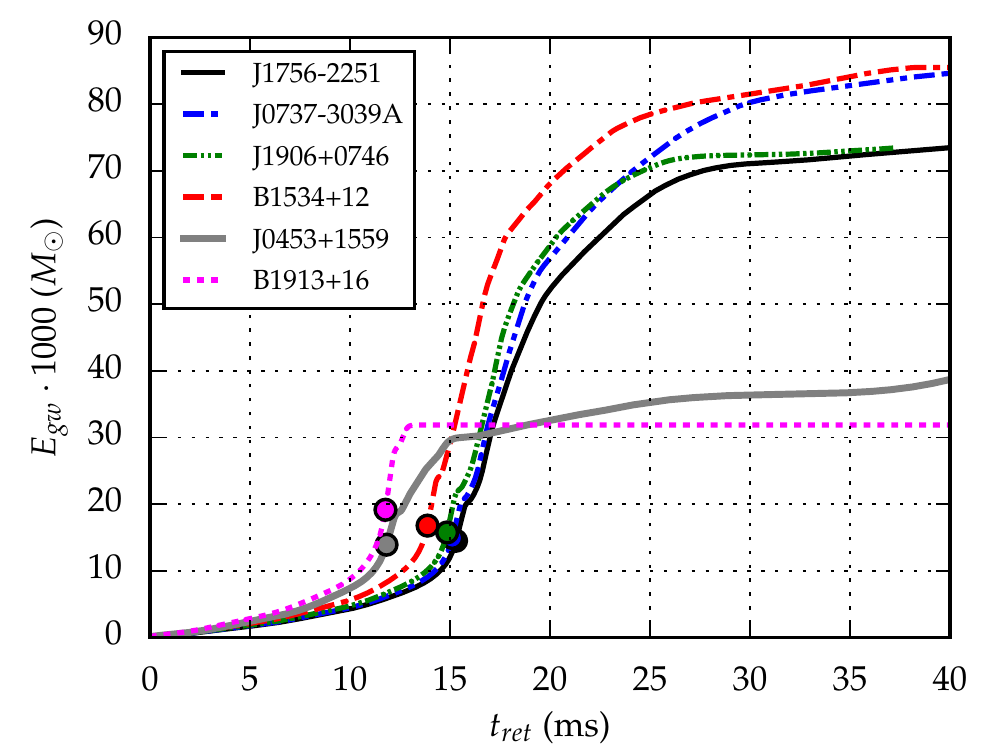}\\
\end{centering}
\vspace{-2mm}
\caption{Evolution of the total energy momentum ($E_\mathrm{gw}$) carried away 
by gravitational radiation in solar mass as a function of the retarded time
$t_\mathrm{ret}$, for the six different BNS models. The colored circles mark the
merger time for each model as indicated in Table~\ref{TAB:energyemitted}.
}
\label{fig:GWemitted}
\end{figure}

\newcommand{\Msun}{$M_\odot$}
\begin{table}
\begin{tabular}{lcccc}
\multirow{3}{*}{Model}   & $E_\mathrm{gw}$    & $E_\mathrm{gw}$     & $E_\mathrm{gw}$    & $t_\mathrm{merger}$ \\
        & (\Msun)    & (\Msun)     & (\Msun)    & (ms)  \\
        & [\ldots-5] &  [-5\ldots5] & [5\ldots20] &       \\
\hline
J1756-2251          & $0.0276$  & $0.0488$  & $0.0189$  & 15.3 \\
J0737-3039A         & $0.0277$  & $0.0528$  & $0.0255$  & 15.1 \\
J1906+0746          & $0.0282$  & $0.0582$  & $0.0148$  & 14.0 \\ 
B1534+12            & $0.0297$  & $0.0597$  & $0.0192$  & 13.9 \\
J0453+1559          & $0.0287$  & $0.0269$  & $0.0060$  & 11.8 \\ 
B1913+16            & $0.0317$  & $0.0276$ & $6\cdot10^{-7}$ & 11.8  \\ 
\end{tabular}
\caption{For each model is reported the total energy 
emitted as gravitational waves from infinite separation up to 
\SI{5}{ms} before merger (second column), from \SI{5}{ms} before merger up 
to \SI{5}{ms} after merger (third column) and from \SI{5}{ms} after merger up to
\SI{20}{ms} after merger (forth column). In the fifth column is reported
the merger time in ms. All these values refer to simulations performed with 
resolution $dx=0.25$ CU ($\simeq 369$ m).}
\label{TAB:energyemitted}
\end{table}

In addition to hydrodynamical quantities, we also analyzed the gravitational waves
generated by such encounters.
Fig.~\ref{fig:GWemitted} shows the total energy momentum ($E_\mathrm{gw}$) that is carried away
by gravitational radiation as a function of the retarded time $t=t_\mathrm{ret}$, 
for all six different BNS models.
As can be seen from this figure, the total energy momentum carried away
by GWs starts increasing only slowly at first, goes through a burst of radiation and then
approaches a more or less steady value in the post-merger region at approximately \SI{20}{ms}
after merger. For this reason, we report for each system the total emitted gravitational energy for
three different time intervals in Table~\ref{TAB:energyemitted}. From infinite separation up to \SI{5}{ms} before merger 
(inspiral phase), from \SI{5}{ms} before merger up to \SI{5}{ms} after merger 
(merger phase) and from \SI{5}{ms} after merger up to \SI{20}{ms} after merger 
(post-merger phase). As can be seen, the total amount of emitted  
energy is not much different for all models in the inspiral phase. The same it is true
for the merger phase, except for the B1913+16 system that promptly 
collapses, forms a BH, and thus reduces GW emission noticeably just after the merger.
Another exception is the more extreme unequal mass ($q = 0.75$) J0453+1559 system, which shows
a noticeably lower emission.
This is in accordance with results in~\cite{DePietri:2015lya},
where it was also observed that the GW luminosity is suppressed
as the mass ratio of the two stars decreases.
In the post-merger phase different models show very different emission strengths,
depending on very different excitation levels of the post-merger modes.

\subsection{Properties of the models collapsing to a black hole}
\label{SEC:BH}

\begin{figure}
  \includegraphics[width=0.45\textwidth]{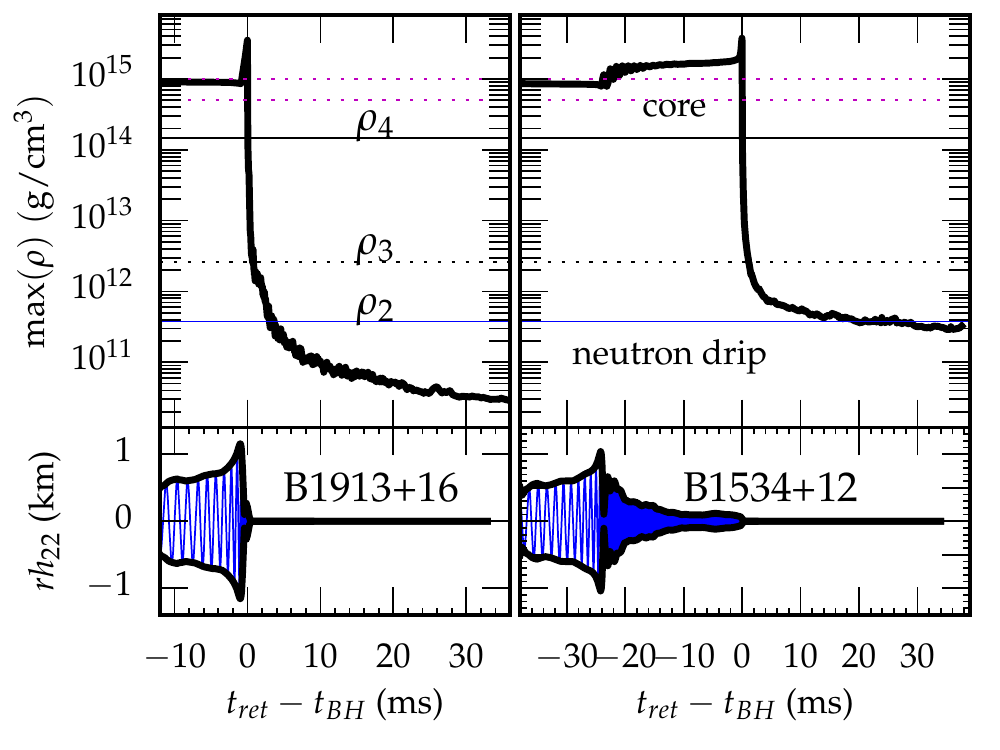}  
\caption{Evolution of the maximum baryonic density as a function of time
for the evolution of systems B1913+16 (left) and B1534+12 (right).
Bottom panel: GW envelope waves emitted by both systems. 
Notice that the maximum density (on the whole numerical grid)
quickly drops below the neutron drip density just after the formation
of the BH ring-down.}
\label{fig:MAXrhoBH}
\end{figure}

As previously noted, of all simulated models, two
collapse to a BH. B1319+16 does so right after the merger (after 
about \SI{0.98}{ms}). On the other hand, B1534+12 does  so only after a short 
hypermassive neutron star (HMNS)
phase, lasting around \SI{24.2}{ms}. In both cases, shortly after BH formation the
maximum baryonic density drops below the neutron drip threshold (see
Fig.~\ref{fig:MAXrhoBH}), and  the whole dynamics of the final disk around and
accreting onto the BH is described by the single-piece of the EOS with
$\Gamma=1.28733$.

\begin{figure}
  \includegraphics[width=0.45\textwidth]{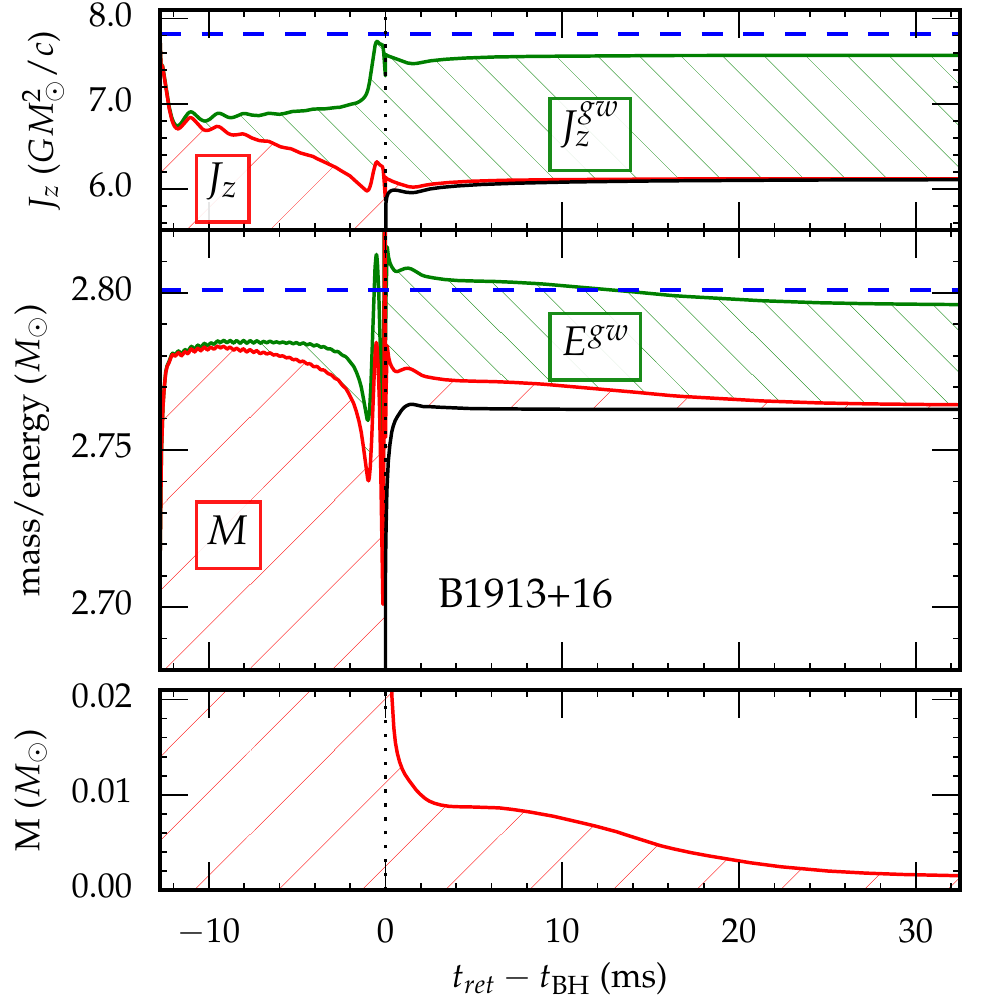}  
\caption{Angular momentum (top-panel) and mass budget (middle-panel) for the evolution 
of model B1913+16. The blue, dashed, horizontal line indicates the 
ADM values of the initial data, as calculated by LORENE. The red-hatched 
area shows the matter contribution, while the green-hatched area shows the 
contributions from the emitted GWs, and finally,
the black-solid line represents the BH contribution. 
The bottom panel shows the total gravitational mass of
the matter present on the numerical grid, zoomed-in to highlight the late-time
accretion.
}
\label{fig:MJbudget}
\end{figure}

For B1913+16, almost
all matter quickly enters the newly formed BH within the first millisecond, 
to arrive at a final BH mass of $M_\mathrm{bh}=2.76$ $M_\odot$ and a dimensionless 
rotational parameter $a=J_\mathrm{bh}/M_\mathrm{bh}^2=0.80$. The measured values (on the numerical
determined apparent horizon) of the mass ($M_\mathrm{bh}$) and angular momentum 
($J_\mathrm{bh}$) do not change, within the numerical error,
past \SI{5}{ms} after BH formation. At the same time, the total 
gravitational mass present in numerical grid (outside of the horizon) is $M=7.5,3.0,1.6$ 
($10^{-3}M_\odot$) \num{10}, \num{20} and \SI{30}{ms} after the formation of the BH, 
respectively. The details of the overall dynamics concerning the 
energy $E_\mathrm{gw}$ and the angular momentum $J_\mathrm{gw}$ carried away by 
gravitational radiation and the BH formation is shown in  
Fig.~\ref{fig:MJbudget}. From there it also apparent that
about \SI{30}{ms} after BH formation the accretion dynamics has stabilized,
within the time-scales considered in these simulations.

\begin{figure}
  \includegraphics[width=0.45\textwidth]{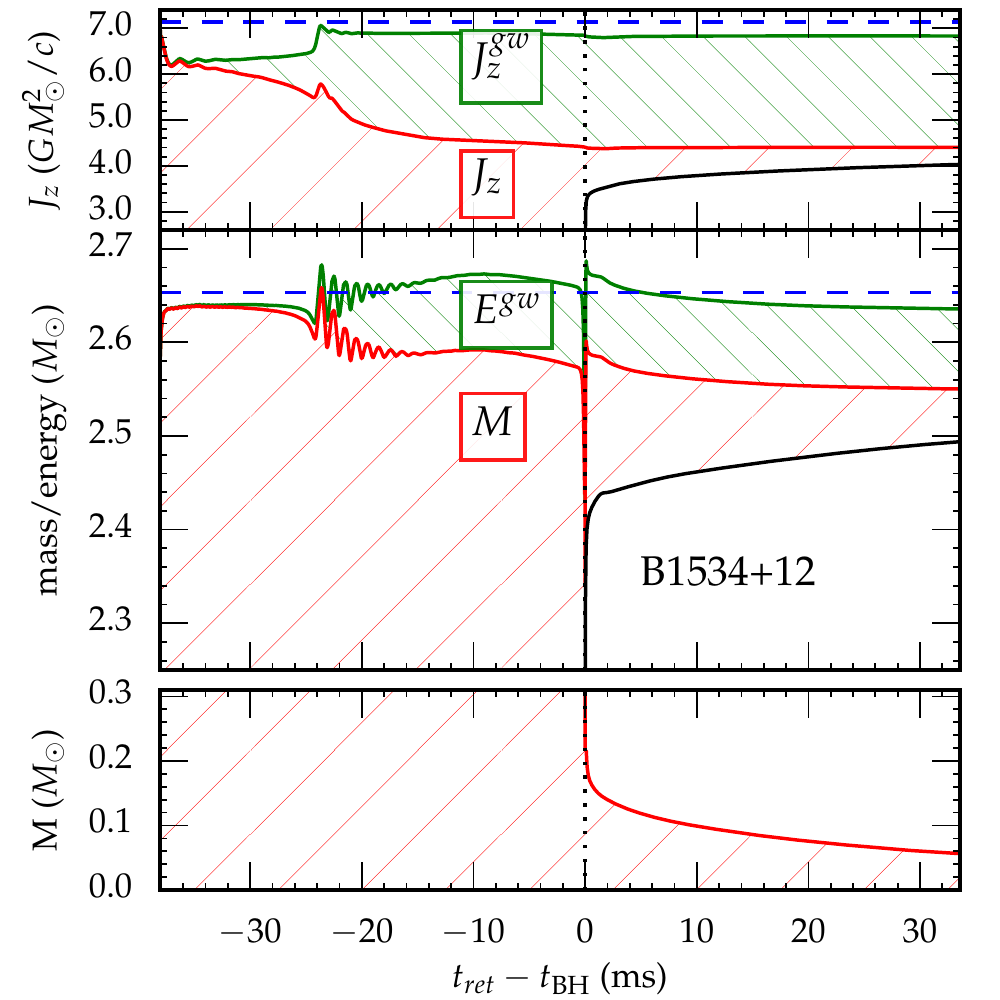}
\caption{Angular momentum (top-panel), mass budget (middle-panel) and total
matter present in the grid (bottom-panel) for the evolution 
of model B1534+12. This figure is similar to Fig.~\ref{fig:MJbudget} 
}
\label{fig:MJbudget2}
\end{figure}

In contrast to B1913+16, B1534+12 does not show a prompt collapse,
as can be seen in a similar Fig.~\ref{fig:MJbudget2}. The BH formation
itself also happens slower in that case. \SI{5}{ms} after BH formation
the BH is characterized by a $M_\mathrm{bh}=2.45$ $M_\odot$ and $a=0.61$, and
these values keep
increasing to their final measured values (\SI{30}{ms} after BH formation)  
of $M=2.49$ $M_\odot$ and $a=0.65$. At that time, the computational grid
still contains matter of total gravitational mass of $M_{disk}=0.06$ $M_\odot$ and
angular momentum $J_{disk}=0.398$ $GM_\odot/c^2$.
Form this information we can deduce that
the final state of the BH will be $M=2.49$ $M_\odot$ and $a=0.68$,
since the amount of gravitational energy and angular 
momentum carried away by gravitational radiation from this moment on 
is negligible. 
The accretion rate $A_r=dM_\mathrm{bh}/dt$ of the remnant BH \SI{35}{ms} after BH
is $\simeq 0.86$ $10^{-3} M_\odot$/ms, is still decreasing but stabilizing.

To summarize, systems B1534+12 and B1913+16 will both collapse to a very fast
rotating BH, characterized by
a dimensionless rotational parameter $a=J_\mathrm{bh}/M_\mathrm{bh}^2$ of
$0.68$ and $0.80$, respectively. However, B1534+12 will show a short-lived
HMNS, while B1913+16 collapses promptly.

\begin{figure*}
\begin{centering}
  \includegraphics[width=0.90\textwidth]{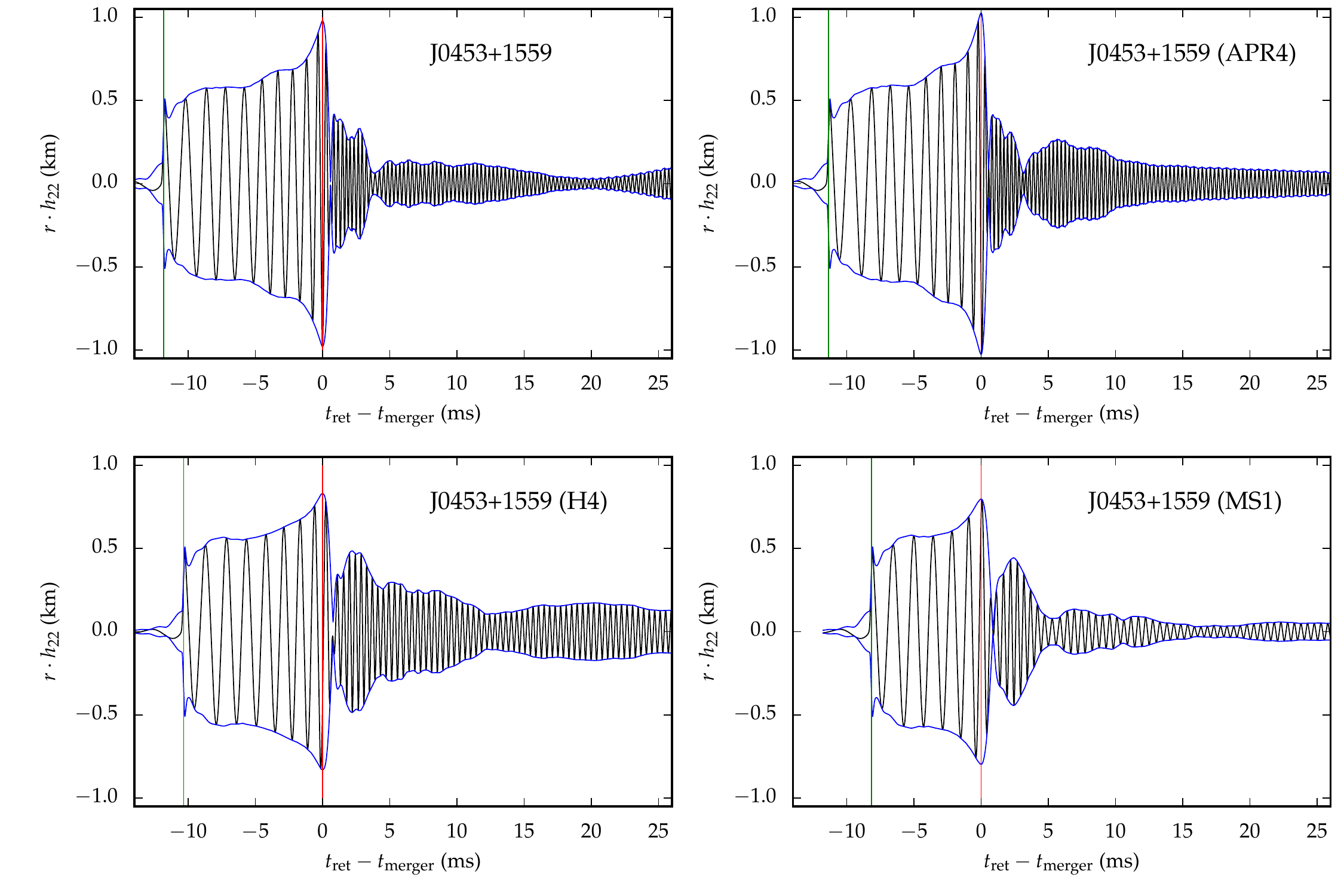}\\
  \includegraphics[width=0.90\textwidth]{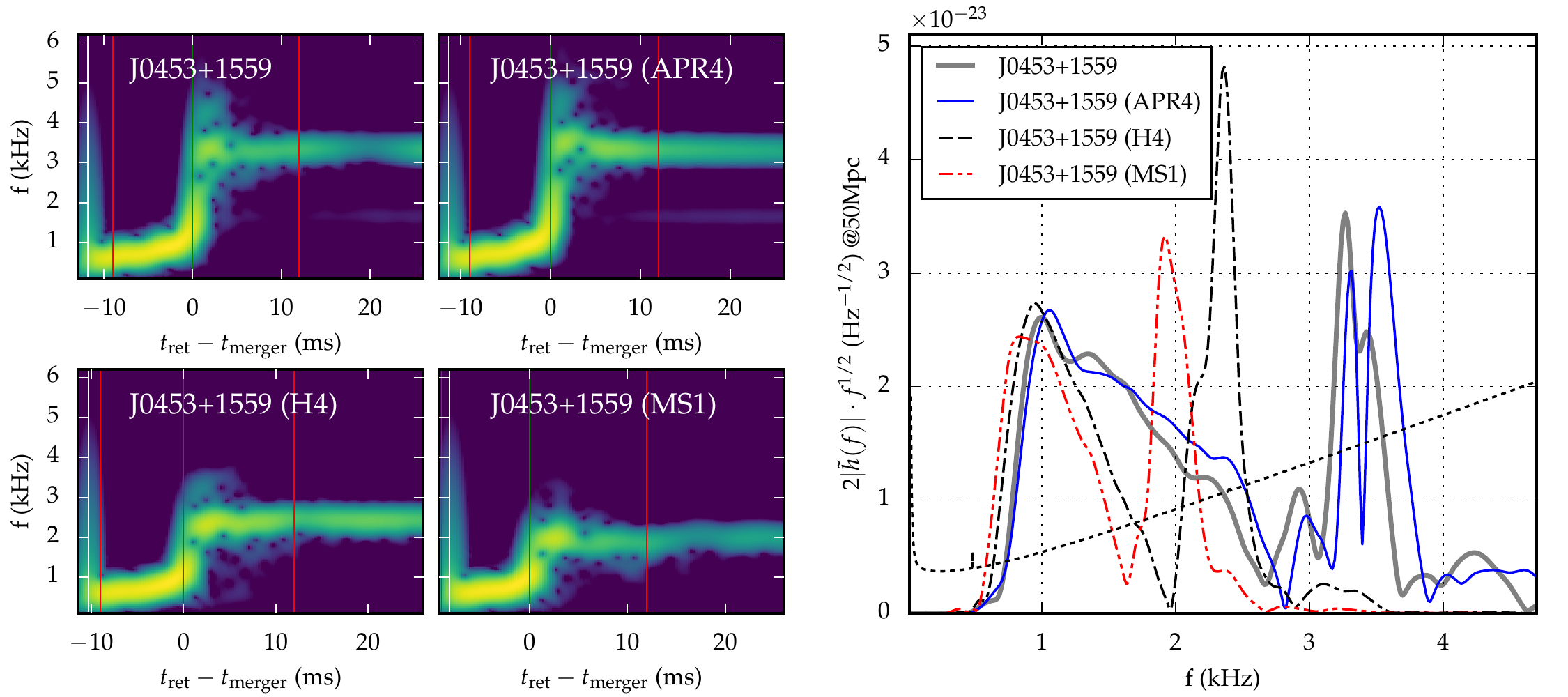}
\end{centering}
\vspace{-2mm}
\caption{Overview of the evolution of model J0453+1559 using different EOS. The meaning of 
plot is the same of Fig.~\ref{fig:all}. One should note the strong suppression
of the side peaks of the main one and the presence of a double peaks structure of the main
in the case of the SLy and APR4 EOS.}
\label{fig:q75}
\end{figure*}

\subsection{Effect of the EOS on the J0453+1559 ($q=0.75$) galactic system}
\label{SEC:q=0.75}

We study the effects of different EOSs on the recently discovered \cite{Martinez:2015mya} BNS 
model associated to the pulsar J0453+1559, which is the most asymmetric known binary system,
characterized by the mass ration of $q=0.75$. Here we analyze, as in \cite{Lehner:2016lxy},
the effect of the EOS on an unequal mass BNS. In particular, we used the three additional
EOSs (see Section~\ref{sec:setup}) APR4, H4 and MS1.  The use of a different EOS changes
the threshold for (following the terminology introduced by~\cite{Baumgarte:1999cq})
a (supra)massive neutron star (SMNS), or a (hyper)massive neutron star (HMNS),
as can be seen in Tab.~\ref{tab:NS_type_thresholds}.
\begin{table}
 \begin{tabular}{lcc}
 \multirow{2}{*}{EOS} & SMNS & HMNS \\
                      & ($M_\odot$) & ($M_\odot$) \\\hline
 SLy & 2.04~(2.42) & 2.41~(2.82) \\
 H4  & 2.01~(2.30) & 2.37~(2.70) \\
 APR4& 2.19~(2.66) & 2.60~(3.09) \\
 MS1 & 2.75~(3.30) & 3.29~(3.90) \\
 \end{tabular}
 \caption{Maximum masses for a SMNS and HMNS for a set of 4 different EOSs. Quantities in
 brackets show the corresponding conserved baryonic mass.}
 \label{tab:NS_type_thresholds}
\end{table}

From this it is clear that for the J0453+1559 system, with a total conserved baryonic mass of
$3.01$ $M_\odot$, the final state does not need to be a BH in the case of the
MS1 EOS. Since its total baryonic mass is less than the maximum mass for a
non-rotating star, it will not even be a SMNS.
On the other hand, the merger remnant will be a SMNS in the case of the APR4 EOS
and a HMNS in the case of the SLy and H4 EOSs, for the same baryonic mass.
  
We evolved J0453+1559 for all the four EOS for about \SI{30}{ms} after merger. None of these
collapsed to a black hole within this time. Fig.\ref{fig:q75}
(similar to Fig.~\ref{fig:all})
shows the amplitude of the $r\cdot h_{22}$ mode (GW signal, top panel) 
as a function of $t - t_\mathrm{merger}$ while on the bottom (right side panel) 
we show the PSD of the effective GW $\bar{h}(f)$ as a function of the frequency in the range 
$[-9,+12]$~ms around merger time. The bottom
left panel shows the spectrogram, using \SI{5}{ms} bins, as a function of time.
Contrary to Fig.~\ref{fig:all} (different binary systems), the simulated GW signal for
J0453+1559 is very different depending on the EOS. In particular, the case of the
H4 EOS shows the strongest post-merger signal corroborated with the biggest peak 
(the dashed black one) in the PSD panel and a frequency $f_p = 2.443$ kHz
(see Table~\ref{TAB:peakFrequencies}. The SLy EOS and APR4 EOS models appear very similar,
although clearly recognizable, while the other two EOSs, namely the H4 EOS and the MS1 EOS
look very different. Differences like these might be useful to infer information about
the EOS from signals measured by interferometers like LIGO/Virgo.

\begin{table}
\begin{tabular}{lccccc} 
\multirow{2}{*}{EOS} & $\tau_0$ & $f_0$ & $f_p$ &$\hat{f}_p$ &$\hat{f}$ \\ 
    & (ms)     & (kHz) & (kHz) &   (kHz)    &   (kHz)  \\ 
\hline
SLy  & 2.54 & 0.998 & 3.331 &  3.268   & 3.429 \\ 
APR4 & 1.83 & 1.136 & 3.310 &  3.310   & 3.520 \\ 
H4   & 2.79 & 0.992 & 2.439 &  2.359   &       \\ 
MS1  & 2.05 & 1.048 & 1.998 &  1.916   &       \\ 
\end{tabular}
\caption{Main peak frequencies and damping times of the post-merger
phase of the simulation of the J0453+1559 system using different choices for the 
EOS, using a resolution of $dx=0.25$ CU. 
$\tau_0$ and $f_0$ are the damping time and the frequency of 
the oscillation of the lapse $\alpha$ between \num{1.5} to \SI{8}{ms} after $t_\mathrm{merger}$. 
The frequency $f_p$ is 
determined by a single-frequency fitting
procedure on the last part of $h_{22}$ signal, 
while the other two $\hat{f}_p$, $\hat{f}$ are derived by 
the analysis of the Fourier spectrum (bottom-right panel of Fig.~\ref{fig:q75}). 
There are other secondary peaks but they are not reported here since 
they are below the LIGO/Virgo design sensitivity.}
\label{TAB:peakFrequenciesEOS}
\end{table}

The post-merger gravitational wave signal shows in all cases a quite complex
evolution, especially for the first few milliseconds. The GW signal is damped,
with the measured frequencies and damping times listed in
Table~\ref{TAB:peakFrequenciesEOS}. These values likely directly depend on
the EOS, as well as the specific merger dynamics, but this dependency cannot
be quantified from the present simulations, and requires further investigations.
It should be noted that in some of these cases the main peak shows a split, and
we denote the additional frequency as $\hat{f}$.
In particular, for the SLy EOS this frequency corresponds to \SI{3.429}{kHz} and for APR4 to 
\SI{3.520}{kHz} (see Table~\ref{TAB:peakFrequenciesEOS} and spectrograms 
on the bottom-left panel of Fig.~\ref{fig:q75}).
The $\hat{f}$ frequency is active during the merger phase for times comparable to the
damping time $\tau_0$ of oscillations of the ADM lapse ($\alpha=(-g^{00})^{-1/2}$)
and the maximum matter density.
This effect is much stronger in the case of the APR4 EOS and is almost not present for the 
EOSs that correspond to less compact stars, namely stars described by the H4 
and MS1 EOS, where the oscillations of the maximum density are smaller.
One should also mention that this split does not show up when the masses of the two stars
are almost equal.

We would also like to note that, for system J0453+1559, the 
GW spectrogram (bottom-left panel) shows a small signal at f=\SI{1.67}{kHz}
starting at \SI{12}{ms} after merger for SLy, and from \SI{10}{ms} after merger
in the case of the APR4 EOS, but not noticeably for the other EOSs.

\begin{figure}
\begin{centering}
  \includegraphics[width=0.45\textwidth]{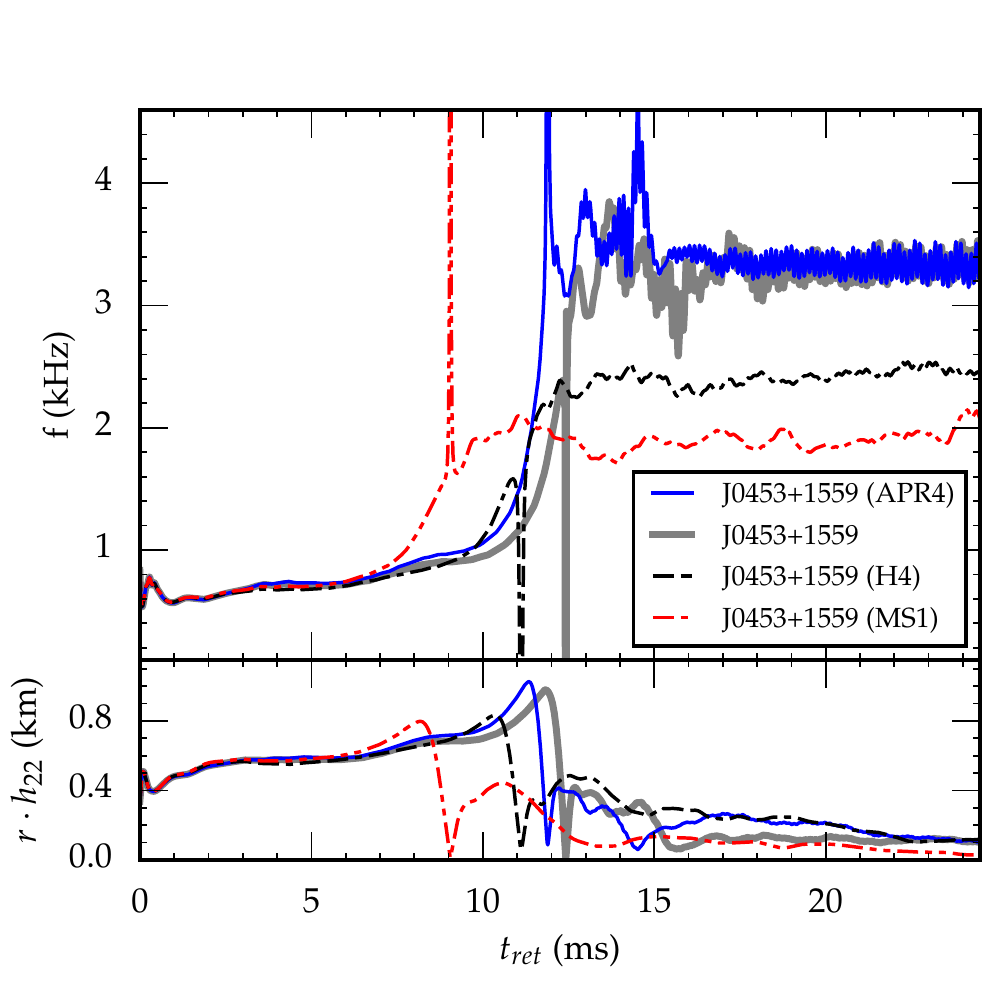}\\
\end{centering}
\vspace{-2mm}
\caption{Evolution of the $22$ instantaneous frequency of the gravitational 
wave signal $f$ (where  $f(t)=d \arg(h_{22}(t))/dt$), as function of the
retarded time ($t_\mathrm{ret}$), for the J0453+1559 system, and
depending on the EOS. The evolution is essentially identical for the
first \SI{5}{ms}, and the effect of the EOS only starts to be visible
after about \SI{7}{ms}.}
\label{fig:freqEOS}
\end{figure}

In order to better understand the dynamics of the GWs emission, we show
the instantaneous frequency of the $h_{22}$ component of the 
GW signal, namely to the quantities $f(t)=d \arg(h_{22}(t))/dt$, in Fig.~\ref{fig:freqEOS}.
We observe almost no difference between evolutions
using different EOS for the first \SI{5}{ms}, while after this
the evolution shows a very strong influence on the EOS. As it can be seen
in Table~\ref{TAB:peakFrequenciesEOS} and
Fig.~\ref{fig:q75}, there is a strong dependence of the frequency of 
the main peaks on the used EOS. The signals from SLy and and APR4
show similarities also here, the signals from the other two EOSs are
clearly different.

\begin{figure}
\begin{centering}
  \includegraphics[width=0.45\textwidth]{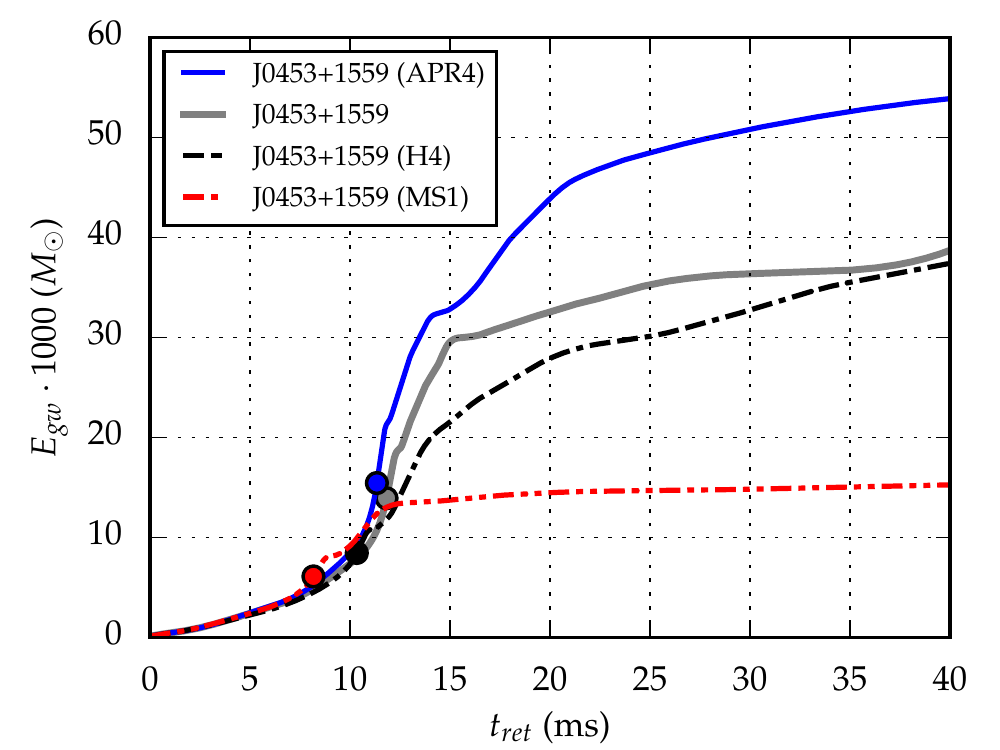}\\
\end{centering}
\vspace{-2mm}
\caption{Evolution of the total energy emitted in gravitational waves
($E_\mathrm{gw}$) for the evolution of J0453+1559, as a function of the retarded
time $r_\mathrm{ret}$ and depending on the used EOS. The energy
is almost identical for the first \SI{5}{ms} where the total 
energies emitted are $2.47$,~$2.41$,~$2.23$, and $2.42$  $10^{-3} M_\odot$ 
for the APR4, SLy, H4 and MS1 EOS, respectively.}
\label{fig:energyEOS}
\end{figure}

\begin{table}
\begin{tabular}{lcccc}
EOS   & $E_\mathrm{gw}$    & $E_\mathrm{gw}$     & $E_\mathrm{gw}$    & $t_\mathrm{merger}$ \\
      & (\Msun)    & (\Msun)     & (\Msun)    & (ms)  \\
      & [\ldots-5] &  [-5\ldots5] & [5\ldots20] &       \\
\hline
SLy    & $0.0287$  & $0.0269$  & $0.0060$  & 11.8 \\ 
APR4   & $0.0295$  & $0.0319$  & $0.0161$  & 11.3   \\
H4     & $0.0276$  & $0.0197$  & $0.0109$  & 10.3   \\
MS1    & $0.0270$  & $0.0125$  & $0.0012$  & 8.17   \\ 
\end{tabular}
\caption{Total energy emitted as gravitational waves for system J0453+1559,
from infinite separation up to 
\SI{5}{ms} before merger (second column), from \SI{5}{ms} before merger up 
to \SI{5}{ms} after merger (third column) and from \SI{5}{ms} after merger up to
\SI{20}{ms} after merger (forth column), for four different EOSs. The fifth column shows
the merger time.}
\label{TAB:energyemittedEOS}
\end{table}

Similarly, and as shown in Fig.~\ref{fig:energyEOS} and Table~\ref{TAB:energyemittedEOS},
while there is almost no dependence of the total energy emitted in GWs
for $5-7$~ms after merger, there are clear differences after that.
The APR4 model shows the highest amount of radiated energy, while the model
using MS1 shows considerably less, and also stops emitting noticeably much
sooner than the other three models.

\section{Conclusions}
\label{sec:conclusions}

We have presented a study of the merger of six different known galactic systems
of BNSs of unequal mass with a mass ratio between $0.75$ and
$0.99$: J1756-2251, J0737-3039A, J1906+0746, B1534+12, J0453+1559 and B1913+16. 
For all of these, we used a semi-realistic, seven segment piece-wise polytropic SLyPP EOS with a 
thermal component of $\Gamma_{th}=1.8$. 
For the most extreme of these systems ($q=0.75$, J0453+1559), we also have 
investigated the effects of different EOSs, namely: APR4, H4, and MS1.

We follow the dynamics of the merger from the late stage of the inspiral
process up to $\sim$\SI{20}{ms} after the system has merged, either to form a
HMNS or a rotating BH.  All simulations have been performed 
using only publicly available, open-source codes such as, the Einstein Toolkit 
code deployed for the dynamical evolution,
and the LORENE code for the generation of the initial models.
All simulations start with the stars at a distance of 
\SI{44.3}{km}, which corresponds to 3-4 orbits before merger, and
all use a resolution of $dx=0.25$~CU.
We followed the late-inspiral phase and post-merger for more than
\SI{20}{ms} (in some cases up to \SI{40}{ms}).

A detailed comparison with the now huge literature 
(see \cite{lrr-2008-8,Lehner:2014asa,Baiotti:2016qnr}) on 
BNS systems is difficult and beyond the scope of the present work. This is particularly 
difficult because in most of the work full details on the numerical setting and a
detailed descriptions of the EOS are lacking. Also, in most of the numerical work the crust 
part of the EOS is not considered or just approximated by a single polytropic EOS. In fact, in
the interpretation of the results one should keep in mind that the final scenario
is strongly dependent on the EOS, i.e., that the post-merger remnant may be 
either a (supra)massive neutron star (SMNS), a (hyper)massive neutron star (HMNS), 
or directly a BH surrounded by an accretion disk.

In all evolutions presented in this work, the amount of baryonic matter
leaving the grid is less than 0.01~$M_\odot$. Using the SLy EOS (in its piece-wise
polytropic representation used here), of the simulated 
systems, J1756-2251 and J0737-3039A
result in a SMNS (very close to the threshold $M_{max}^{SLy}=2.82\ M_\odot$ to
be HMNS), while all the other correspond to HMNSs. Only two of the HMNSs
collapse to a BH within the simulated time (\SI{35}{ms}); one right after the 
merger (B1913+16, Hulse-Taylor system), and the second after a short 
hyper massive NS phase formation of \SI{25}{ms} (B1534+12). 
These cases have some resemblance to the already analyzed cases 
in~\cite{DePietri:2015lya}, namely models SLy16vs16 and SLy15vs15, 
respectively. Both models result in a BH surrounded by an accretion disk,
but with a disk of very much different mass. The final dimensionless 
rotational parameter of the BH ($a=J_\mathrm{bh}/M_\mathrm{bh}^2$) is $0.80$ for
B1913+16, while the expected final value for B1534+12 is $0.68$.

We also considered the evolution of J0453+1559, using also three different EOSs.
The remnant just after the merger is a HMNS for SLy and H4, while
it is a SMNS in the case of APR4, and in the case of M1 the system is even below the
non-rotating maximum mass. Nevertheless, none of these models collapses to a BH
within the first \SI{35}{ms} after merger. We have also
shown that the inspiral phase of the merger is not very
sensible to the EOS. Dominant effects should be related to the eccentricity, as
shown in~\cite{Maione:2016zqz}, but not studied here.
Tidal effects during this phase may in principle be detected by future
LIGO/Virgo/KAGRA detectors and will depend on eccentricity. Thus, it is
crucial to reduce the orbital eccentricity in simulations to maximize the precision 
with which we can extract neutron-star parameters and constrain the 
neutron star EOS from GW observations. This task 
has already been considered in~\cite{Kyutoku:2014yba}, and we will leave 
the task to find the appropriate method to generate
the (still-lacking) public initial data with no eccentricity to future work.

Most of the energy in GWs is emitted during the merger (from \SI{5}{ms} before
to \SI{5}{ms} after it). The GW signal during the inspiral is essentially
identical for all models. On the other hand, the total amount of energy emitted 
in the post-merger (from \SI{5}{ms} to \SI{20}{ms} after merger) 
is almost of the same size as the total energy emitted during the whole 
inspiral phase (from infinite separation to \SI{5}{ms} before merger).
We note that for the more extreme unequal mass case (J0453+1559)
the total energy emitted is much reduced, in agreement with the 
results of Ref.~\cite{DePietri:2015lya}, where it was shown that the 
GW luminosity is suppressed as the mass ratio of the two stars decreases.
There is also the notable exception of B1913+16, which promptly
collapses to a BH and shortly after stops emitting GWs. This is similar to
binary black hole mergers, which would make it more difficult to discriminate
this case from a BBH merger.

Finally, we would like to remark that the merger and post-merger phases
show a very complicated GW pattern, stemming likely from very complex
excitation mode dynamics. Because of this, we expect it to be unlikely
that semi-analytic templates could be 
assembled to represent the full GW signal from the inspiral to the post-merger
phase.
The only common characteristic (beside the one from inspiral phase up to five 
\si{ms} before merger) seems to be that there is a final and sustained phase 
in the late post-merger dynamics in which there is only a dominating
mode in the 22-component.

\acknowledgments

This project greatly benefited from the availability of public software that
enabled us to conduct all simulations, namely ``LORENE'' and the ``Einstein
Toolkit''. We do express our gratitude the many people that contributed to
their realization.

This work would have not been possible without the support of the 
SUMA INFN project that provided the financial support of the work of AF
and the computer resources of the CINECA ``GALILEO'' HPC Machine, where most
of the simulations were performed. Other computational resources were provided 
by he Louisiana Optical Network Initiative (QB2, allocations loni\_hyrel, loni\_numrel,
and loni\_cactus), and by the LSU HPC facilities (SuperMike II, allocations
hpc\_hyrel15 and hpc\_numrel).
FL is directly supported by, and this project heavily used infrastructure 
developed using support from the National Science Foundation in the USA 
(Grants No. 1212401, No. 1212426, No. 1212433, No. 1212460).
Partial support from INFN ``Iniziativa Specifica TEONGRAV'' and by the ``NewCompStar'', 
COST Action MP1304, are kindly acknowledged.

\end{document}